\documentclass[submitted]{IEEEtran}
\usepackage{graphicx}
\usepackage{overpic}
\usepackage{url}

\usepackage{cite}
\usepackage{color}
\usepackage{upgreek}
\usepackage{textcomp,mathcomp}
\usepackage[colorlinks=true,linkcolor=blue,urlcolor=blue,citecolor=blue]{hyperref}
\usepackage[colorinlistoftodos]{todonotes}
\usepackage{makecell}
\usepackage{multirow}
\usepackage{multicol}
\usepackage{soul, color, xcolor}
\PassOptionsToPackage{prologue,dvipsnames}{xcolor}
\definecolor{c1}{HTML}{00ffff}
\definecolor{c2}{HTML}{c0c0c0}
\definecolor{c3}{HTML}{00ff00}
\definecolor{c4}{HTML}{ffc0cb}

\begin{document}

\title{\bf A Bi-polar Current Source with High \\ Short-Term Stability for Tsinghua \\ Tabletop Kibble Balance}
\author{Kang Ma, Xiaohu Liu, Wei Zhao, Songling Huang, Shisong Li$^{\dagger}$

\thanks{K. Ma, X. Liu, W. Zhao, S. Huang, and S. Li are with the Department of Electrical Engineering, Tsinghua University, Beijing 100084, China. W. Zhao is also with the Yangtze Delta Region Institute of Tsinghua University, Jiaxing, Zhejiang 314006, China.}
\thanks{This work was supported by the National Key Research and Development Program of China under Grant 2022YFF0708600 and the National Natural Science Foundation of China under Grant 52377011.}
\thanks{$^\dagger$Email: shisongli@tsinghua.edu.cn}}

\maketitle

\begin{abstract}
A high-precision current source, capable of supporting weighing measurements with a relative uncertainty at the $10^{-9}$ level, is essential for Kibble balance experiments. However, most current sources utilized in Kibble balances to date are homemade and not commercially available.
In this paper, we introduce a digital-feedback, two-stage current source designed for the Tsinghua tabletop Kibble balance, relying solely on commercially available sources and voltmeters. A high-resolution, small-range current source is employed to digitally compensate for current output fluctuations from a large-range current source. Experimental tests show the proposal can offer an easy realization of a current source with nA/A stability to support Kibble balance measurements.
\end{abstract}


\section{Introduction}\label{introduction}
\IEEEPARstart{T}{he} Kibble balance, originally known as the watt balance~\cite{Kibble1976}, is one of the main approaches to realize the unit of mass, the kilogram, in the new International System of Units (SI)~\cite{CIPM}. The Kibble balance can virtually link mechanical power to electrical power through two measurement phases, i.e., the weighing phase and the velocity phase.
During the weighing phase, a coil excited by a DC current $I$ is placed in a magnetic field, and the electromagnetic force generated by the current-carrying coil is counterbalanced by the weight of a test mass, written as
\begin{equation}
BlI = mg,
\label{eq:weighing}
\end{equation}
where $B$ is the magnetic flux density at the coil position, $l$ is the coil wire length {($Bl$ is also known as the geometrical factor)}, $m$ is the test mass, and $g$ is the local gravitational acceleration.
In the velocity phase, the geometrical factor $Bl$ is calibrated by moving the coil in the same magnetic field and measuring the ratio of the induced voltage on the coil terminals, $U$, to the coil's moving velocity $v$, i.e.,
\begin{equation}
Bl = \frac{U}{v}.
\label{eq:velo}
\end{equation}

Substituting (\ref{eq:weighing}) into (\ref{eq:velo}), a virtual power balancing equation, $mgv = UI$, is obtained, and hence the test mass is calibrated as
\begin{equation}
m = \frac{UI}{gv}.
\label{eq:mass}
\end{equation}

On the right side of (\ref{eq:mass}), {$U$ and $I$} are measured against quantum electrical standards~\cite{zimmerman2010quantum,li2020mu}, and {$g$ and $v$} are measured by interferometer-based instruments and traced to frequency and length standards. Finally, the mass is linked to the Planck constant, $h$, by expressing the quantities on the right side in quantum form, as detailed in~\cite{haddad2016bridging}.

{More than} a dozen Kibble balance experiments are ongoing at National Metrology Institutes (NMIs)~\cite{NRC,NIST,METAS,LNE,NIM,KRISS,MSL,UME,PTB}, as well as the International {Bureau of Weights and Measures (BIPM)}~\cite{BIPM} and other metrology laboratories~\cite{THU2023}. The most accurate Kibble balance can calibrate a kilogram level mass with a relative uncertainty of approximately $1\times10^{-8}$. {To obtain such} accuracy, each quantity on the right side of (\ref{eq:mass}) should be measured with uncertainty at the level of $10^{-9}$. 

In the context of Kibble balances, achieving {an extremely} stable and highly accurate current source is feasible, as demonstrated in various studies~\cite{NPL, NISTCS}, though it remains a challenging task. Several research groups have reported on the design and performance of current sources for weighing measurements.
In the NPL/NRC Kibble balance, two 16-bit {digital to analog converters (DACs)} with a gain ratio of -2000:1 are summed to serve as the input for amplification~\cite{NPL}. The first-stage source supplies a $\pm16$\,mA ($\pm$500\,g) output with a resolution of 0.5\,$\upmu$A (16\,mg). The second-stage source outputs $\pm8\,\upmu$A ($\pm$250\,mg) with a resolution of 0.25\,nA (8$\upmu$g). The NIST-3 and NIST-4 systems employ a similar scheme~\cite{NISTCS}, utilizing two 20-bit DACs combined with a ratio of 1000:1. It achieves a noise level of approximately 100\,pA/$\sqrt{\mathrm{Hz}}$ at 1\,Hz and a short-term stability of 0.1\,(nA/mA)/hour~\cite{haddad2016invited}. The METAS Kibble balance uses a custom-designed, low-noise current source for weighing measurements~\cite{metas2022}. The weighing current is $\pm 6.5$\,mA with a noise level of about $40$\,nA/A (20\,$\upmu$g) over 120\,s. The current source used by the LNE Kibble balance group can output $\pm 5$\,mA. It employs the sum of two 16-bit DACs, with resolutions of 152\,nA and 380\,pA, respectively~\cite{LNE}. The stability of this current source, aided by a real-time control loop and compensation from the Josephson voltage standard, achieves an Allan deviation of 1\,nA/A over 30\,s of measurement. The Joule balance group presented a two-loop feedback current source~\cite{wang2014250}, achieving a relative stability of approximately 200\,nA/A in a 30-minute test at 250\,mA.

The design and application of precision current sources extend beyond the Kibble balance field, finding relevance in laser diode drivers, magnetometers, ultra-low-noise measurements, and magnetic field generators. 
To meet the low-noise requirements of laser diodes, Christopher \textit{et al.} propose a current source with a microprocessing unit to control the current set point digitally~\cite{Christopher-2008}. This design achieves a noise level of approximately 10\,nA at 1\,Hz with an output current of 74.5\,mA and maintains a low temperature coefficient of 1.7\,ppm/$^\circ$C. Based on the Hall-Libbrecht current driver~\cite{Hall-1993}, Daylin \textit{et al.} developed an ultra low-noise current source with a range of $\pm$50\,mA and a noise level of 2\,pA at 1\,Hz~\cite{Daylin-2013}. Similarly, Matthew \textit{et al.} propose a low-noise current source with a range of 500\,mA, achieving a noise level below 2\,nA at 1\,Hz by improving hardware and introducing a regulator block~\cite{Taubman-2011, Taubman-2013}. Xia \textit{et al.} introduce a composite topology to enhance Howland current source, whose long-term stability is less than 220\,ppm with a range of 0\~487.3\,mA~\cite{Xia-2022}.
In the domain of ultra low-noise measurements, Carmine \textit{et al.} propose to use a high stability battery instead of a solid-state voltage reference to reduce $1/f$ noise~\cite{Carmine-1998}. The designed low-noise current source is with a range of 50\,mA and a noise level of 10\,pA at 1\,Hz. Scandurra \textit{et al.} {developed} a high-impedance, programmable current source using a low-noise Junction Field Effect Transistor (JFET) and a programmable floating voltage source, achieving a noise level of 6\,pA at 1\,Hz at 1.8\,mA within a 50\,mA range~\cite{Scandurra-2014}.
For high-sensitivity magnetometers, a high dynamic range and ultra low-noise current source is essential. Wang \textit{et al.} {designed} a programmable current source using two {DACs}, with a range extending to $\pm$202\,mA. {This current source achieves a noise level of 30\,pA at 1\,Hz at 150\,mA and a stability down to 30\,nA/A}~\cite{Wangyanzhang-2022}.
In the field of magnetic field generation, high stability current sources are crucial. Kyu-Tae Kim \textit{et al.} propose a feedback scheme current source based on voltage references, {achieving a stability down to 20\,nA/A}~\cite{Kyu-2005}. To further improve stability at milliampere levels, the research group of Physikalisch-Technische Bundesanstalt (PTB) {developed} a method using external references, achieving stability down to 1\,nA/A at 50\,mA~\cite{PTBcurrent-2019}.

As evident from the literature review, precise current sources are predominantly customized or homemade by enhancing the hardware, particularly for applications requiring both a wide range (up to $\pm20$\,mA) and low noise or high resolution. This raises an interesting question: Can commercial source modules be directly utilized to achieve a precision current source for Kibble balance measurements? Some Kibble balance groups have attempted to use commercial current sources, as documented in studies such as~\cite{UME, PTB}. Table \ref{tab:commerical} lists several precision commercial current source models along with their typical performance specifications. It is evident that even the best state-of-the-art commercial current sources, such as the Keysight B2961A/B2962A, struggle to fully meet the stringent requirements for high-precision Kibble balance measurements.

\begin{table}[tp!]
    \centering
     \caption{List of some high-precision commercially available current sources. }
    \label{tab:commerical}
    \begin{tabular}{c|c|c|c}
    \hline\hline
         Model 	& Range	& Resolution 	&Peak noise@1Hz \\
          	& /mA	& /$\upmu$A & /$\upmu$A \\
         \hline 
        Keithley 6220/6221	& 20	& 1	& 2 \\
        Keysight B2961A/B2962A	& 10	& 0.01	& 0.1 \\
        Keithley 2410	& 20	& 0.5	& 0.2 \\
        Keithley 2400/20/25/30	& 10	& 0.5	& 0.05 \\
        Yokogawa GS200	& 10	& 0.1	& 0.2 \\
        \hline
    \end{tabular}
\end{table}

Unlike the traditional approach of designing and customizing a high-precision homemade current source, we present an alternative method to achieve a precision current source for Kibble balance measurements by combining two commercially available sources with different ranges. In this method, a small-range, high-resolution source compensates for the fluctuations of a large-range, low-resolution current source using full-digital feedback controlled by LabVIEW. This proposed current source will be employed in the Tsinghua tabletop Kibble balance experiment~\cite{THU2023}.

The paper is organized as follows: In Section \ref{sec:02}, using the Tsinghua tabletop Kibble balance as an example, we discuss the general requirements for a current source in Kibble balance measurements. Section \ref{principle} introduces the principle and realization of the digital-feedback two-stage current source. Section \ref{main} presents the experimental tests of the proposed current source. Finally, Section \ref{conclusion} provides the conclusion.

\section{General requirements for a Kibble balance current source}
\label{sec:02}
In a Kibble balance, achieving a kilogram mass standard requires an optimal $Bl$ value in the range of a few hundred Tm to minimize overall measurement uncertainty, as discussed in \cite{schlamminger2013design, li2022irony}. Considering mass-on and mass-off measurements \cite{Stephan16}, the required current is given by:
\begin{equation}
    I = \pm \frac{mg}{2Bl},
    \label{eq:range}
\end{equation}
where the positive and negative signs correspond to the mass-on and mass-off measurements during the weighing phase, respectively. 

Eq. (\ref{eq:range}) defines the output range of the current source. For calibrating a kilogram mass, the typical current ranges from a few mA to about 20\,mA. In the Tsinghua tabletop Kibble balance, $Bl\approx400$\,Tm and the currents required for mass-on and mass-off are $\pm 12.5$\,mA for calibrating a 1\,kg mass. 

\begin{figure}[tp!]
    \centering
    \includegraphics[width=0.5\textwidth]{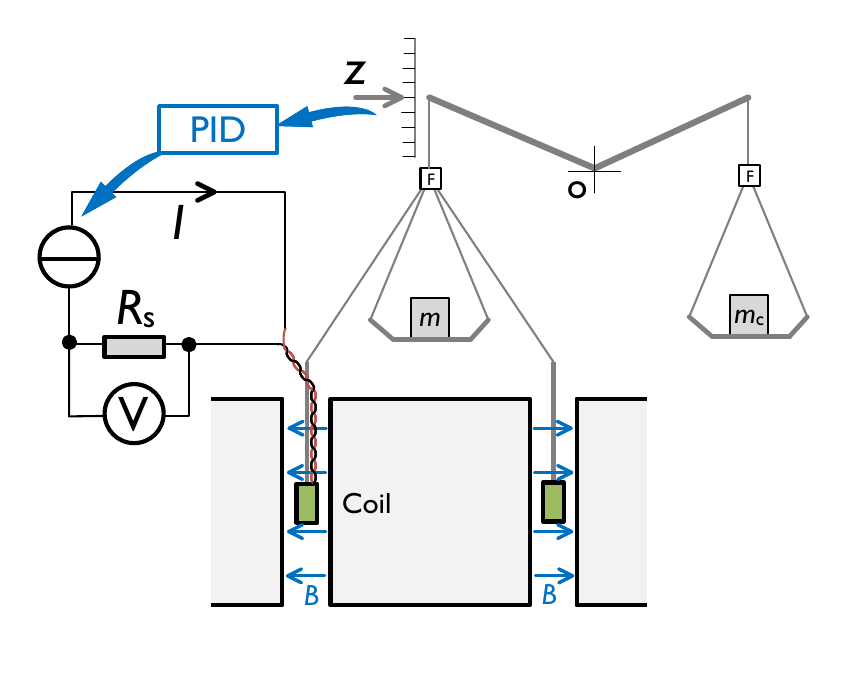}
    \caption{The servo control in a typical Kibble balance. O is the rotation center of the balance beam or a flexure hinge. $m$ is the mass to be calibrated and $m_c$ the counterweight mass. The current carrying coil is placed in the magnetic field ($B$) and electromagnetic force $BlI$ is counterbalanced by the weighing of the mass. The residual force is reflected in the coil position $z$, and the position difference, $\Delta z=z-z_\mathrm{set}$ is modulated by a PID feedback. Finally, the current is adjusted to minimize $\Delta z$. The stabilized current is measured by the voltage drop on a standard resistor $R_\mathrm{s}$. }
    \label{fig:weighingloop}
\end{figure}

For a conventional Kibble balance, there is current running through the coil only during the weighing measurement. As shown in Fig.\ref{fig:weighingloop}, the current source is integrated as part of the force measurement feedback loop and varies to ensure that the residual force is null or that the weighing position remains fixed, as seen in \cite{NIST,NRC}. In the stable state of the control loop, the total force acting on the pivot is balanced, i.e.,
\begin{equation}
    mg-BlI+\Delta F=0, 
\end{equation}
where $\Delta F$ is the residual force from the mechanical system. Either the drift of $Bl$ (mainly thermal and current-related effects~\cite{li18,li2022,li2022irony}) or $\Delta F$ (such as the relaxation of the flexure~\cite{NIST2024flexures}, slow deformation of the sensitive mechanical parts, etc.) can yield a change of the current, therefore, in this case, the current $I$ is not constant during the weighing measurement. {To simplify the nomenclature,} this working state of the current source is named in the following text the 'constant force' (CF) operation scheme.

{In} the CF scheme, $I$ is not constant and varies with drifts from the balance or the magnet system, so long-term stability is not an important target for the current source design. Instead, the resolution and noise level become significant. Obtaining a resolution of $10^{-9}$ with a single DAC {requires more than 30 bits}, which is not commercially available. Like the NIST and NRC Kibble balances~\cite{NPL, NISTCS}, a classic solution is to sum two DACs with different resolutions. If both DACs are $N$ bits and their output ratio is $K$, the current output then can be written as
\begin{equation}
    I=\frac{U_\mathrm{ref}}{R}\left(\alpha \frac{1}{2^N}+\beta \frac{1}{2^N}\frac{1}{K}\right),
\end{equation}
where $\alpha$ and $\beta$ are integer numbers within the DAC range, i.e. $-(2^N-1)\leq\alpha,\beta\leq 2^N-1$; $K$ is the output ratio of two DACs, and $|K|>1$; $U_\mathrm{ref}/R$ is the voltage-current conversion coefficient. For example: $N=20$, $U_\mathrm{ref}/R=20$\,mA, $K=1000$, and the current required for the weighing measurement is $\pm 12.5$\,mA. In this case, the current resolution offered by the main DAC is {19\,nA ($0.02/2^{20}$\,A), or $1.5\times10^{-6}$ relatively (19\,nA/12.5\,mA),} {while the fine DAC yields a resolution of 19\,pA or $1.5\times10^{-9}$,} reaching the required current resolution for the weighing measurement. 

Note that such a two-stage source has a high precision only in servo control loops, because the stability of the open-loop output is limited by the noise of the first stage source. With the closed loop, the noise of the first stage can be compensated in the bandwidth and hence can give {a highly stable} or low-noise output. The final output noise level depends on how fast the servo-control loop is. In principle, the faster the better. Ref. \cite{Stephan16} pointed out that the speed of the servo control loop for the weighing measurement is limited due to two factors: the mechanical resonance (sub Hz to tens of Hz) and the time constant of the circuit, $\tau$. For the Tsinghua Kibble balance, a bifilar coil is used for weighing and velocity measurements. {For a bifilar coil with multi-turns, the stay capacitance between wires, compared to the coil inductance ($L=1.5$\,H) and the resistance ($R_{\rm{t}}\approx 500\,\Omega$), is negligible}~\cite{Stephan16}{. Hence, the time constant of the circuit can be estimated as $\tau \approx L/R_{\rm{t}}\approx3$\,ms.} In this case, the time-constant limit is not comparable to the mechanical resonance. In addition, considering the integration time of {the DVM} during the measurement, a bandwidth within 100\,Hz can well satisfy the servo control requirement. 

There is a second setup for the current source, i.e., during mass-on and mass-off measurements, the current through the coil is {kept constant}. {This is defined as} the 'constant current' (CC) operation scheme. In this case, the current source is running on an electrical feedback loop, and the residual force, $BlI-mg$, is usually not zero. This scheme is typically used in Kibble balances equipped with a commercial weighing cell, such as those in \cite{METAS,BIPM,NIM}. The weighing cell can read out the residual force from its internal feedback loop. Note that in the CC scheme, not only the high resolution but also high stability is required for the current source. {In the one-mode method especially,} where the current is running during both weighing and velocity measurements, unstable current during the velocity measurement phase could introduce undesired induced voltage and hence bias on the $Bl$ measurement, known as the coil-current effect~\cite{li17}. For example, the coil magnetic flux linkage due to the current is $IL$ (for a bifilar coil, the mutual inductance $M$ of two coils equals to the inductance of each single coil, $L$), and hence the additional $Bl$ related is written as
\begin{equation}
\Delta (Bl)=\frac{\partial(IL)}{\partial t}\frac{\partial t}{\partial z}=I\frac{\partial L}{\partial z}+\frac{L}{v}\frac{\partial I}{\partial t},
\label{eq:currenteffect}
\end{equation}
where $v=\partial z/\partial t$ is the coil moving velocity. Studies of the first term on the right side of (\ref{eq:currenteffect}) can ensure the effect on the $Bl$ measurement is well compensated or corrected~\cite{nistmag,BIPMmag2017,li18}{, while} the effect caused by the second term mainly depends on $\partial I/\partial t$. The ideal case is letting $\partial I/\partial t=0$ during the measurement. In this case, the current can not have a considerable drift or fluctuation during the measurement, and hence, good stability of the current is necessary. 

It can be seen from the above analysis that the CC operation scheme is more crucial for the current design compared to the CF scheme. In principle, a current source for the CC scheme should be able to work with the CF scheme. 
There are some other factors, such as the impedance to the ground, calibrations of key electrical components, etc, that need to be considered for a Kibble balance current source design. However, those issues can always be tested and fixed using known measures, and therefore are not focused on in this paper. 

\section{Design and principle of the digital-feedback two-stage current source}
\label{principle}

\begin{figure*}[tp!]
    \centering
    \includegraphics[width=0.9\textwidth]{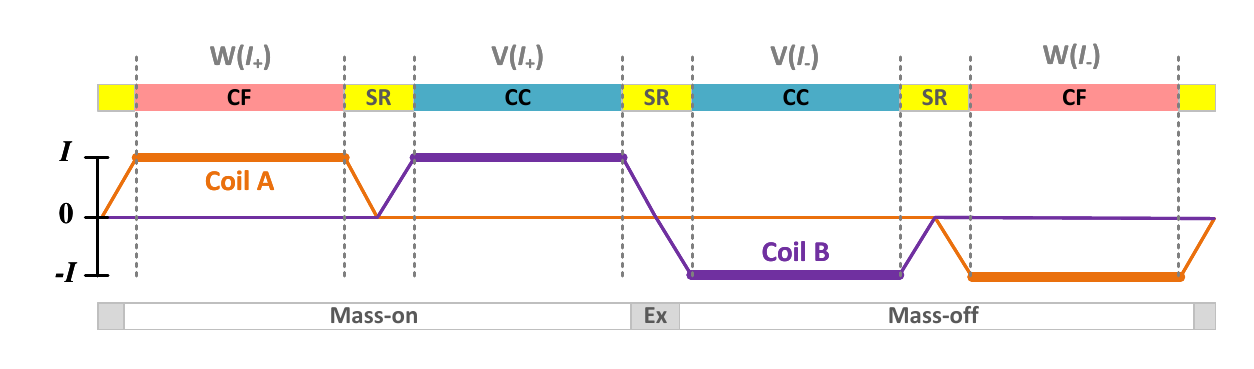}
    \caption{A typical measurement sequence for OMTP Kibble balance (one measurement period is shown). CC and CF denote respectively the constant current scheme and the constant force scheme. SR presents the switch and current ramping and Ex means the mass exchange. W and V are weighing and velocity measurement phases. }
    \label{fig: current source scheme}
\end{figure*}

The Tsinghua tabletop Kibble balance employs the one-mode, two-phase (OMTP) measurement scheme. A bifilar coil, comprising two parallel wound independent coils (coil A and coil B), is utilized. For example, considering coil A as the measurement coil, Fig.\ref{fig: current source scheme} illustrates a typical OMTP measurement sequence within a single period. A complete measurement encompasses four stages: mass-on weighing $W(I_+)$, mass-on velocity $V(I_+)$, mass-off velocity $V(I_-)$, and mass-off weighing $W(I_-)$.
During the $W(I_+)$ stage, the current source supplies a current $I$ to coil A in the CF scheme, while coil B remains open-circuited. The subsequent phase is the $V(I_+)$ stage, wherein the current source switches to supply a current $I$ to coil B in the CC scheme, while coil A is open for induced voltage measurement. Following this, the test mass is unloaded, and the $V(I_-)$ stage is conducted. The current through coil B becomes $-I$, and coil A remains open for $U/v$ measurement. The final phase is the $W(I_-)$ stage, wherein the current source provides a current of $-I$ for weighing in coil A, with coil B open-circuited.
Each stage duration ranges from several minutes to over ten minutes. The use of symmetrical currents aids in suppressing significant systematic effects, such as the current effect \cite{BIPM}.

\begin{figure}[htbp]
    \centering
    \includegraphics[width=0.5\textwidth]{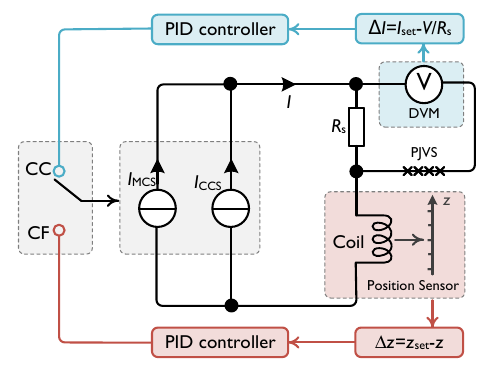}
    \caption{Principle of DTCS under different schemes. In which, $z_{\rm{set}}$ is the $z$-axis coordinate of the set position of the coil. $z$ is the measured position. $I_{\rm{set}}$ is the requirement current value under CC scheme.}
    \label{fig: principle of DTCS}
\end{figure}

{Accordingly,} the current source for the Tsinghua tabletop Kibble balance must operate in both CF and CC schemes. Therefore, it requires not only high resolution but also high stability. Currently, commercial current sources cannot meet the stringent requirements under conditions of mA-level current output. Customizing a high-precision current source is challenging for most groups. Here we propose an easily implementable method to achieve an ultra-precision current source relying solely on commercial current sources and voltmeters. The design integrates two commercially available sources using a digital-feedback, and hence in the following text is named digital-feedback, two-stage current source (DTCS). The overall design of the proposed DTCS is depicted in Fig.\ref{fig: principle of DTCS}. The main current source (MCS) and the compensating current source (CCS) are connected in parallel to provide current to a well-calibrated, high-stability resistor ($R_{\rm{s}}$) and the coil, i.e.
\begin{equation}
    I=I_\mathrm{MCS}+I_\mathrm{CCS}.
\end{equation}
To be able to compensate for fluctuations of MCS, the output range of CCS should be significantly larger than the resolution of MCS. Meanwhile, the resolution of CCS, compared to the required current for the weighing measurement, {should be on the order of $10^{-9}$ relatively}. 

When the DTCS operates in the CC scheme, the output of the MCS is set to the required current value. A high-resolution digital voltmeter (DVM) and a programmable Josephson quantum voltage standard (PJVS) are used to measure the voltage on the sampling resistor $R_\mathrm{s}$ terminals. Consequently, the current within the circuit can be determined with high precision. The difference between this measured current value and the required current value, denoted as $\Delta I$, serves as the input for a proportional-integral-derivative (PID) controller.
To mitigate the impact of random fluctuations, a low-pass filter or simply averaging of samplings over a specific period is applied to $\Delta I$ and the filtered signal is used as the input to the PID controller. The PID controller then adjusts the output of the CCS to stabilize the fluctuations of the MCS. The current generation strategy could be simple: First, ramp and fix the MCS to the set current within its maximum resolution, and then turn on the CCS feedback loop to ensure the fine digits are stable. 

When the DTCS operates in the CF scheme, the position of the coil is monitored by a laser interferometer or other precision measuring instruments. The difference between the $z$-axis coordinates of the set position and the measured position is input to the PID controller, which provides the required current value. Depending on the range of the MCS and CCS, the required current is divided into two parts: the main part handled by the MCS, $I_\mathrm{MCS}$ and the minor part handled by the CCS, $I_\mathrm{CCS}$. Here we show an example of the current distribution for MCS and CCS. {The resolution of the first stage current source is 0.5\,$\upmu$A in the range of 20\,mA, and the range of the second stage is 2\,$\upmu$A.} With a set current value $I$ (unit: A), the outputs of MCS and CCS can be calculated as follows, i.e.
\begin{eqnarray}
    I_\mathrm{MCS}&=&\mathrm{ROUND}(I\times10^6)/10^6,\\
    I_\mathrm{CCS}&=&[I\times10^6-\mathrm{ROUND}(I\times10^6)]/10^6,
\end{eqnarray}
where ROUND($x$) is a function giving the nearest integer to $x$. For example, when $I=12.5012455$\,mA is required, the above formula yields $I_\mathrm{MCS}=12.501$\,mA, and $I_\mathrm{CCS}=0.2455\,\upmu$A.   

In both schemes, the current sources, DVM, and coil position measuring instrument are controlled by a LabVIEW program, which implements the digital PID controller.
To meet the stringent requirements of the Kibble balance, the range of the MCS should reach 20\,mA, and the resolution of the MCS should be as high as possible. The range of the CCS should cover the resolution of the MCS, usually in the $\upmu$A level with a resolution below 1\,nA. Many existing commercial current sources can meet such requirements. In the Tsinghua Kibble balance configuration, the MCS is a Keithley 2410 (20\,mA range, 0.5\,$\upmu$A resolution) and the second stage is a Keithley 6221 (2\,$\upmu$A range, 0.1\,nA resolution). 

When the PJVS is used, the DVM measures only the residual voltage, and in this case, the primary requirement for the DVM is high resolution, which can be easily met using commercial nanovoltmeters. However, without the PJVS, the DVM should have the highest possible number of digits for precision CC feedback. In the Tsinghua Kibble balance system, a Keysight 3458A is employed.
Additionally, to ensure the accuracy of the measurement current, the sampling resistor should be well-calibrated with excellent stability. The stability during measurement should reach $1\times10^{-9}$. The optimal parameters for the PID controller should be determined experimentally under varying conditions.

\section{Experimental tests of DTCS}\label{main}

\begin{figure}
    \centering
    \includegraphics[width=0.5\textwidth]{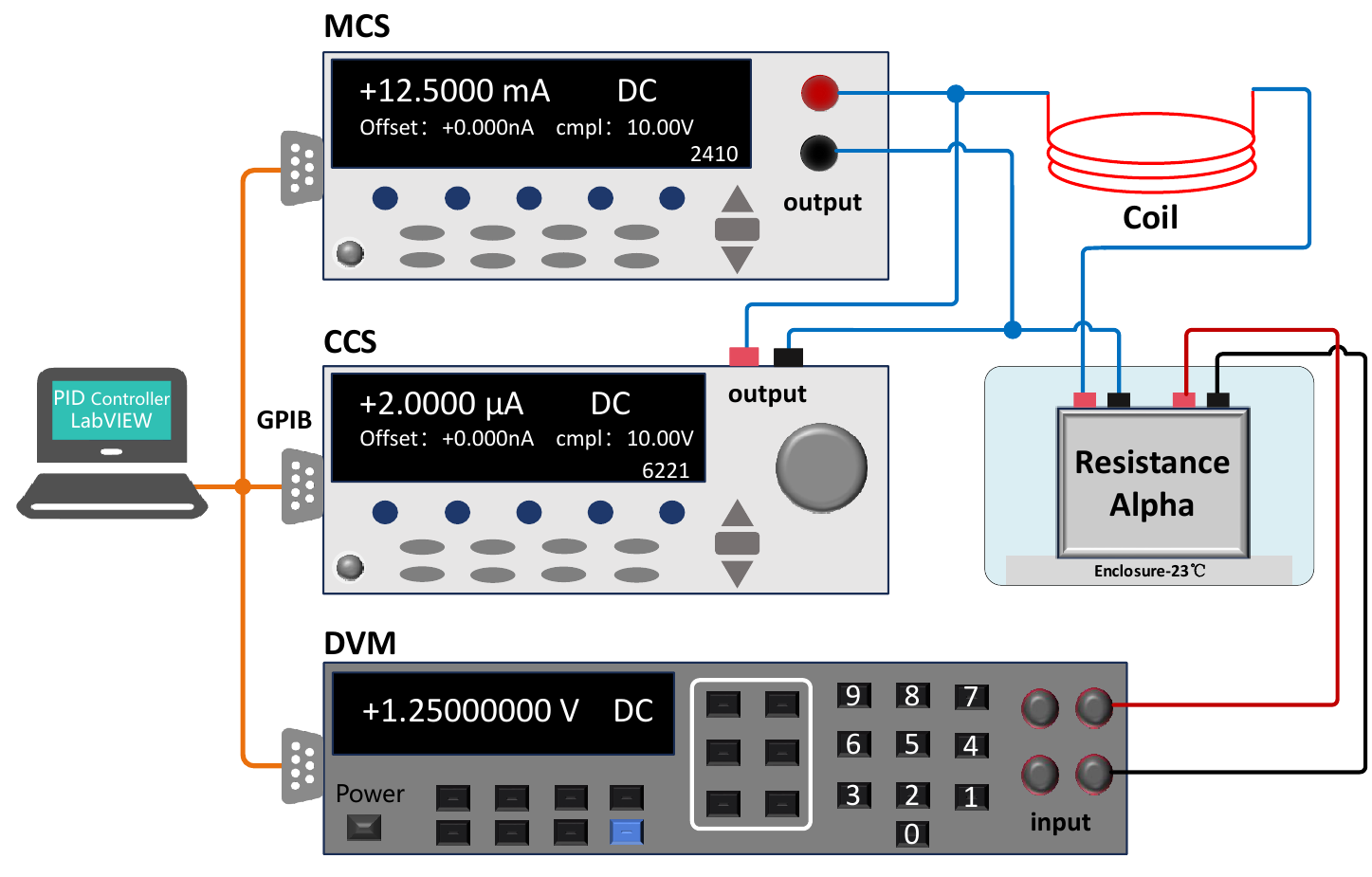}
    \caption{Experimental setup of the test. The MCS and CCS used in the configuration are respectively a Keithley 2410 and a Keithley 6221. The DVM is a 3458A and the resistance is a 100\,$\Omega$ high-stable standard resistor from Alpha Electronics (HRU-100). }
    \label{fig:instruments connections}
\end{figure}

According to Section \ref{sec:02}, the stability requirements for the current source in the CC scheme are more stringent than those in the CF scheme. Therefore, the experiments in this section are conducted under the CC scheme. {The experimental setup is shown in Fig.}\ref{fig:instruments connections}. {The current sources configuration is the same as mentioned in the above section.} {The voltage drop on a standard resistor, $R_\mathrm{s}$ (Alpha-HRU-100), is used as the servo feedback signal. The standard resistor is periodically calibrated against the quantum Hall resistance standard at the National Institute of Metrology (NIM, China), and the latest calibration yields $R_\mathrm{s}=100.00001534\,\Omega$ with a relative uncertainty of $5\times10^{-9}$. A periodic calibration of $R_\mathrm{s}$ allows a track of the resistance drift over time and by making small corrections, the accuracy of the resistance used in the measurement can be ensured at the $10^{-9}$ level.} During the experiments, the resistor operates in an {enclosure} maintained at a temperature of 23$\tccentigrade$ with a fluctuation of $\pm$1\,mK. The stability of $R_{\rm{s}}$ can reach up to {$\pm$0.05\,$\upmu{\rm{\Omega}}/{\rm{\Omega}}$/year}, and the temperature coefficient of $R_{\rm{s}}$ is {$\pm$0.05\,$\upmu{\rm{\Omega}}/{\rm{\Omega}}$/°C.} Thus, the drift of $R_{\rm{s}}$ during the experiment is negligible.
An 8.5-digit multimeter, Keysight 3458A, is utilized as the DVM to measure the voltage across $R_{\rm{s}}$. The range of the 3458A is set to 10\,V, with the autozero function enabled. The number of digits (NDIG) is set to $8{\raise0.5ex\hbox{$\scriptstyle 1$}\kern-0.1em/\kern-0.15em\lower0.25ex\hbox{$\scriptstyle 2$}}$, and the number of power line cycles (NPLC) is set to 10. A bifilar coil, which will be applied to the Tsinghua tabletop Kibble balance, is integrated into the circuit. The coil has 1360 turns and a resistance of approximately 400\,$\Omega$.
Before the input to the PID controller, a moving-average filter with window length of 2.5\,s is used to smooth the current difference signal $\Delta I=I-I_{\rm{set}}$.

\begin{figure}
    \centering
    \includegraphics[width=0.5\textwidth]{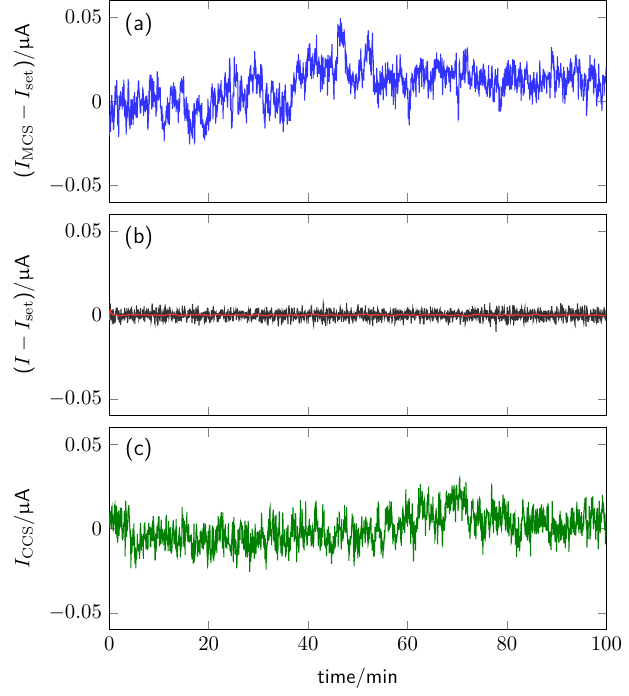}
    \caption{The time domain signal of measurement current. The duration of the test is 100\,min.  (a) is the MCS. (b) is the DTCS. (c) is the compensation current in DTCS. Here $I_\mathrm{set}=12.5$\,mA. The red line in (b) is the 100s-average value of measurement current in the circuit, and its fluctuation is about 1.5\,nA.}
    \label{fig:timedomain for MCS and DTCS}
\end{figure}

\begin{figure}[tp!]
    \centering
    \includegraphics[width=0.5\textwidth]{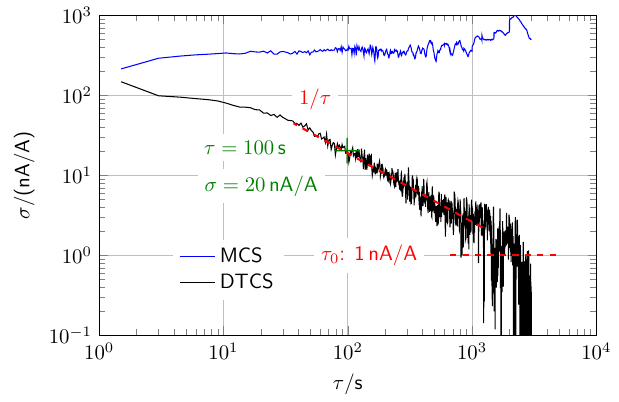}
    \caption{The Allen deviation of current measurement for MCS and DTCS. {$\sigma$ is the Allen deviation and $\tau$ is integration time.}}
    \label{fig:Allen for MCS and DTCS}
\end{figure}

As noted in Section \ref{sec:02}, the required current for the Tsinghua tabletop Kibble balance is 12.5\,mA. Therefore, the experiments are initially conducted at 12.5\,mA. When the MCS operates independently, the measured current is shown in Fig.\ref{fig:timedomain for MCS and DTCS}(a). Due to the limitation of MCS resolution, the peak-to-peak fluctuation of the measured current is approximately 0.1\,$\upmu$A. {In this paper, the Allen deviation is utilized to characterize the stability of the current. Note that as the result is tested repeatable under the same parameter setup, to be concise, the following plots show only one of the test results.} Accordingly, when the MCS operates independently, the Allen deviation of measurement current is depicted in Fig.\ref{fig:Allen for MCS and DTCS}.  When the integration time is less than 100\,s, the stability is around 250\,nA/A. However, when the integration time exceeds 100\,s, the Allan deviation increases with increasing integration time, indicating that the MCS has poor long-term stability.

For the DTCS, the proportional, differential, and integral parameters are  initially set to 2.5, 0, and 0.2, respectively. The loop time of the PID controller is 500\,ms, mainly due to the DVM measurement and readout time when a high number of digits is set. {After filtering using a moving average (window width of 2.5\,s)}, the measured current in the circuit is shown in Fig.\ref{fig:timedomain for MCS and DTCS}(b). Compared with Fig.\ref{fig:timedomain for MCS and DTCS}(a), the long-term stability of the current after compensation is significantly improved. Meanwhile, the compensation current remains within 0.1\,$\upmu$A, which is within the range of the CCS.
Under the above PID controller parameters, the stability of the DTCS is shown in Fig. \ref{fig:Allen for MCS and DTCS}. When the integration time is 100\,s, {the stability reaches down to $2\times 10^{-8}$.} Compared to the MCS, the stability has been improved by more than a magnitude. Furthermore, with increasing integration time, the stability of the DTCS can reach {down to 1\,nA/A at $\tau\approx30$\,mins.}

\begin{figure}[tp!]
    \centering
    \includegraphics[width=0.5\textwidth]{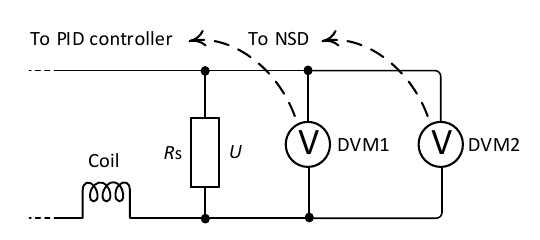}
    \caption{Experimental setup for NSD measurement. DVM1 and DVM2 are used to measure the voltage across the sampling resistor, simultaneously. The measurement result of DVM1 is used by the PID controller for adjusting the output of CCS. The measurement result of DVM2 is used to calculate the NSD. For the low-frequency domain (below 1\,Hz), the settings of 3458A are the same as that of DVM1, and the sampling time is 100\,min. For the frequency higher than 1\,Hz, the sampling rate is 500\,Hz. The continuous sampling time is 10\,s.}
    \label{fig:measurement for NSD}
\end{figure}

\begin{figure}
    \centering
    \includegraphics[width=0.5\textwidth]{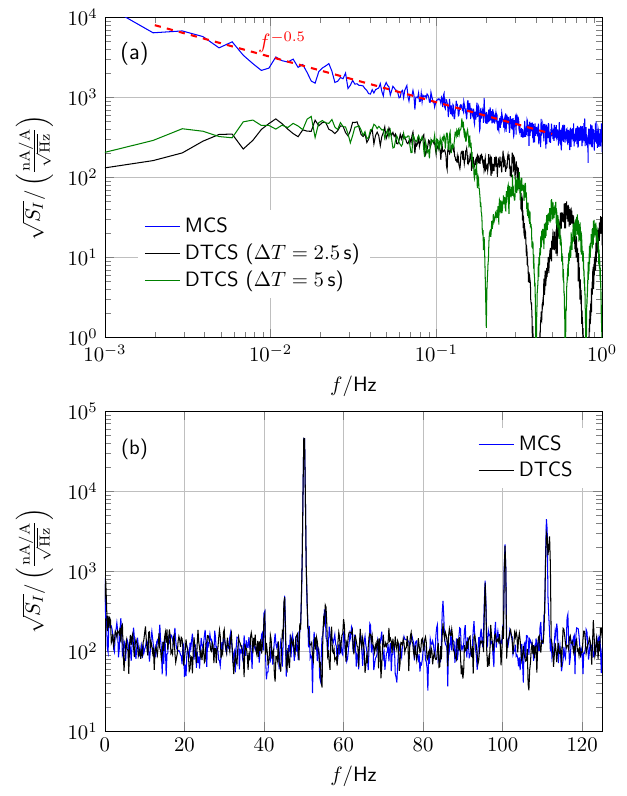}
    \caption{The {NSD} of MCS and DTCS. (a) and (b) respectively shows the spectrum in the range of 1\,Hz and the range of 125\,Hz.}
    \label{fig:results of NSD}
\end{figure}

In Kibble balance experiments, it is essential to consider not only the stability of the current but also the current noise in the frequency domain. Therefore, the frequency characteristics of the signals of both the MCS and DTCS are studied. To obtain the {noise spectral density (NSD)} of the current source, an additional 3458A multimeter is connected in parallel with the resistor, as shown in Fig.~\ref{fig:measurement for NSD}. The results of the NSD measurement are presented in Fig.~\ref{fig:results of NSD}.
When the frequency is below 1\,Hz, the noise density of the DTCS is significantly lower than that of the MCS, indicating that the $1/f$ noise of the MCS has been effectively suppressed by CCS compensation. For frequencies higher than 1\,Hz, the noise density of the DTCS is approximately equal to that of the MCS. Notably, due to the power supply system, the noise density at 50\,Hz and its harmonics are significant.

In Kibble balance experiments, each weighing phase lasts for several minutes, and the average current value is used to calculate the weighing geometrical factor, $Bl$. High-frequency noise attenuates rapidly with increased integration time. Therefore, noise with a frequency higher than 1\,Hz has little impact on the weighing measurement. According to Fig.~\ref{fig:Allen for MCS and DTCS}, when the integration time exceeds 200\,s, {the stability of the DTCS current source can reach $10^{-9}$ level,} meeting the high-precision requirements of the weighing measurement.

It is interesting to note the presence of two notches at frequencies of 0.4\,Hz and 0.8\,Hz on the black curve in Fig.~\ref{fig:results of NSD}(a). These notches result from the 2.5\,s moving-average filter applied to the error signal in the servo control. This filter effectively suppresses components that are inversely proportional to the length of the chosen moving-average window, $\Delta T$ (here $f=1/\Delta T= 0.4$\,Hz), as well as its harmonics ($2f=0.8$\,Hz in this case). To confirm this conclusion, the 5\,s moving-average filter is applied to the error signal in the servo control. The noise density spectrum is shown in the green curve of Fig.~\ref{fig:results of NSD}(a). Four notches at 0.2\,Hz, 0.4\,Hz, 0.6\,Hz, 0.8\,Hz are obtained. As evidenced by the above experiments, the window length, $\Delta T$ can be adjusted by experimenters. By stabilizing the servo loop using a different set of PID parameters, different notches can be designed.

\begin{figure*}[h!]
    \centering
    \includegraphics[width=0.95\textwidth]{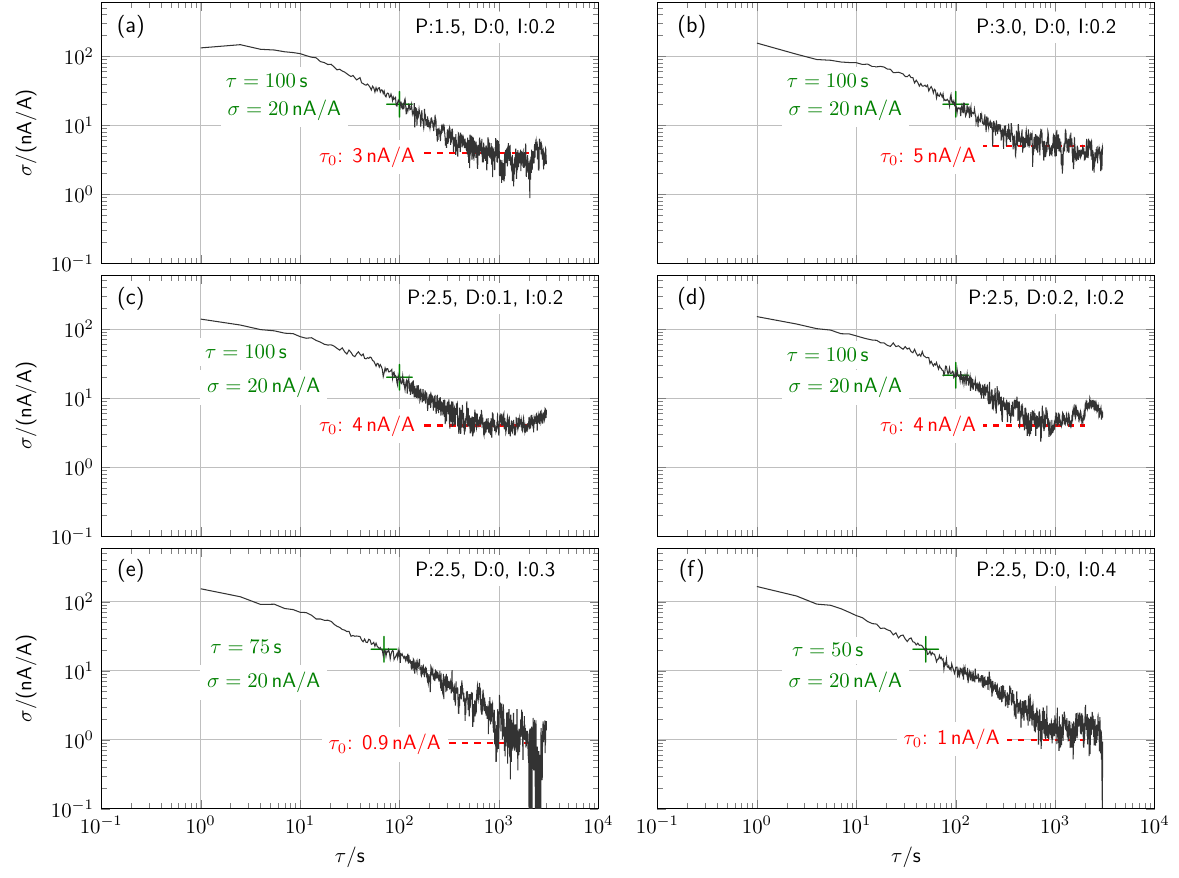}
    \caption{The stability of the DTCS under different PID controller parameters. The reference proportional, differential and integration parameter are 2.5, 0 and 0.2, respectively. In (a) and (b), the proportional parameter is changed. In (c) and (d), the differential parameter is changed. In (e) and (f), the integration parameter is changed.}
    \label{fig:different pid}
\end{figure*}

Since the OMTP measurement scheme is employed, reducing the noise level during the velocity measurement is an important task. The typical configuration of the induced voltage measurement uses three DVMs, e.g., 3458A multimeters, and triggers each DVM in a flat sequence to avoid data omission~\cite{haddad2016invited,BIPM,LNE}. The length of the integration time for the DVM is mainly determined by the spectrum of the induced voltage. Here we need to ensure the noise introduced by the current source, following the coupling path in (\ref{eq:currenteffect}), is as low as possible. The measurement of electrical noise level at different frequencies in Fig.~\ref{fig:results of NSD} offers some useful information for choosing an appropriate integration time for the induced voltage measurement. As shown in Fig.~\ref{fig:results of NSD}, the major noise peaks are the power frequency and harmonics, and hence choosing an integer NPLC value for the DVM can suppress the noise from the current source. Selective notch filters can also be considered to further remove undesired noise peeks, such as the one at 110\,Hz. 

As observed from the principle of the DTCS, the performance of the PID controller significantly impacts the stability of the current source. Here, the stability of the current source under different PID parameters is examined. The results are shown in Fig.~\ref{fig:different pid} for an output current of 12.5\,mA. Firstly, from Fig.\ref{fig:different pid}(a), (b) and Fig.\ref{fig:Allen for MCS and DTCS}, it is evident that when the proportional parameter is either too large or too small, the stability of the DTCS current source degrades. The proportional term primarily influences the response speed to the error and should have an optimal value. 
By comparing the best stabilization achieved, the proportional parameter is fixed at 2.5. 
According to Fig.~\ref{fig:different pid}(c), (d) and Fig.\ref{fig:Allen for MCS and DTCS}, a non-zero differential parameter degrades the stability of the DTCS, and hence the differential parameter is set to 0.
As shown in Fig.~\ref{fig:different pid}(e), (f) and Fig.\ref{fig:Allen for MCS and DTCS}, an appropriate increase in the integration parameter improves the stability of the DTCS. It is noteworthy that a larger integration parameter reduces the integration time required to achieve specific stability. For instance, when the integration parameter is set to 0.4, the integration time to achieve stability of 20\,nA/A is 50\,s, greatly accelerating the servo stabilization. In this study, the integration parameter is fixed at 0.4.

\begin{figure*}[h!]
    \centering
    \includegraphics[width=0.95\textwidth]{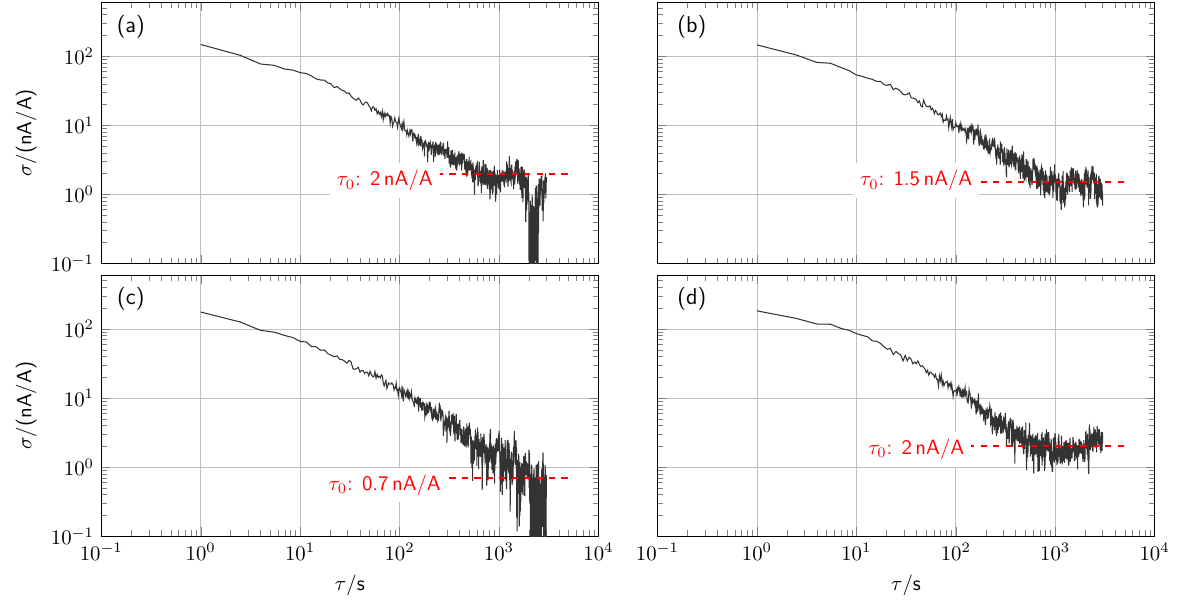}
    \caption{The stability of the DTCS under different currents: (a) for 7.5\,mA, (b) for 10\,mA, (c) for 15\,mA and (d) for 19\,mA. The corresponding PID controller parameters used are: P: 2.5, D: 0, I: 0.4.}
    \label{fig:different current}
\end{figure*}

Typically, the coil current in Kibble balance experiments varies from mA level to a maximum of 20\,mA. {At the end of the test}, the stability of DTCS under different currents is studied. The results are shown in Fig.\ref{fig:different current}. For all cases, the stability of the DFTS current source can reach down to nA/A level when the same PID parameters (P: 2.5, D: 0, I: 0.4) are used.

\section{Conclusion}
\label{conclusion}

This paper demonstrates a DTCS for Kibble balance measurements. The DTCS achieves compensation for current source output fluctuations using a digital controller, relying solely on commercial current sources and voltmeters. Experimental tests show that the proposed DTCS can achieve nA/A stability when the output varies from several mA to 20\,mA, meeting the precision requirements for typical Kibble balance weighing measurements. 
Experimental results confirm that the DTCS can be used in both constant force and constant current operation schemes. The proposed method offers an easy-to-implement alternative to the traditional approach of customizing a high-precision current source, providing a practical and effective solution for achieving high-stability current in Kibble balance measurements.

Looking forward, there is still work to be done. The PJVS system has not yet been integrated into the system, and the precision of the DVM is currently a major limitation for improving the proposed DTCS. Enhancing the update rate of the DTCS servo control is another issue worth investigating and by accelerating the residual voltage measurement, the loop time can be significantly reduced.

\section*{Acknowledgement}
The authors would like to thank Dr. Nong Wang from the Beijing Institute of Control Engineering and Dr. Yunfeng Lu from the National Institute of Metrology (NIM) for valuable discussions. 


\begin{thebibliography}{10}

\bibitem{Kibble1976}
B.~P. Kibble, ``A measurement of the gyromagnetic ratio of the proton by the strong field method,'' in \emph{Atomic masses and fundamental constants 5}. Springer, 1976, pp. 545--551.

\bibitem{CIPM}
Resolution of the 26th CGPM conference, Bureau des Poids et Mesures, Sèvres, France, 2018.

\bibitem{zimmerman2010quantum}
N.~M. Zimmerman, ``Quantum electrical standards,'' \emph{Phys. Today}, vol.~63, no.~8, pp. 68--69, Aug. 2010.

\bibitem{li2020mu}
S.~Li, Q.~Wang, W.~Zhao, and S.~Huang, ``From $\mu_0$ to $e$: A survey of major impacts for electrical measurements in recent SI revision,'' \emph{IEEE Trans. Instrum. Meas.}, vol.~69, no.~9, pp. 5956--5965, Jul. 2020.

\bibitem{haddad2016bridging}
D.~Haddad, F.~Seifert, L.~S. Chao, {\it et al}, ``Bridging classical and quantum mechanics,'' \emph{Metrologia}, vol.~53, no.~5, pp. A83--A85, Sep. 2016.

\bibitem{NRC}
B.~M. Wood, C.~A. Sanchez, R.~G. Green, and J.~O. Liard, ``A summary of the Planck constant determinations using the NRC Kibble balance,'' \emph{Metrologia}, vol.~54, no.~3, pp. 399--409, May. 2017.

\bibitem{NIST}
D.~Haddad, F.~Seifert, L.~S. Chao, {\it et al}, ``Measurement of the Planck constant at the National Institute of Standards and Technology from 2015 to 2017,'' \emph{Metrologia}, vol.~54, no.~5, pp. 633--641, Jul. 2017.

\bibitem{METAS}
H.~Baumann, A.~Eichenberger, F.~Cosandier, {\it et al}, ``Design of the new METAS watt balance experiment Mark II,'' \emph{Metrologia}, vol.~50, no.~3, pp. 235--242, May. 2013.

\bibitem{LNE}
M.~Thomas, D.~Ziane, P.~Pinot, {\it et al}, ``A determination of the Planck constant using the LNE Kibble balance in air,'' \emph{Metrologia}, vol.~54, no.~4, pp. 468--480, Jun. 2017.

\bibitem{NIM}
Z.~Li, Z.~Zhang, Y.~Lu, {\it et al}, ``The first determination of the Planck constant with the joule balance NIM-2,'' \emph{Metrologia}, vol.~54, no.~5, pp. 763--774, Sep. 2017.

\bibitem{KRISS}
D.~Kim, B.-C. Woo, K.-C. Lee, {\it et al}, ``Design of the \textsc{KRISS} watt balance,'' \emph{Metrologia}, vol.~51, no.~2, pp. S96--S100, Mar. 2014.

\bibitem{MSL}
C.~M. Sutton and M.~T. Clarkson, ``A magnet system for the \textsc{MSL} watt balance,'' \emph{Metrologia}, vol.~51, no.~2, pp. S101--S106, Mar. 2014.

\bibitem{UME}
H.~Ahmedov, N.~B. A{\c{s}}k{\i}n, B.~Korutlu, and R.~Orhan, ``Preliminary Planck constant measurements via UME oscillating magnet Kibble balance,'' \emph{Metrologia}, vol.~55, no.~3, pp. 326--333, Apr. 2018.

\bibitem{PTB}
C.~Rothleitner, J.~Schleichert, N.~Rogge, {\it et al}, ``The Planck- balance -- using a fixed value of the Planck constant to calibrate E1/E2-weights,'' \emph{Meas. Sci. Technol.}, vol.~29, no.~7, p. 074003, May. 2018.

\bibitem{BIPM}
H.~Fang, F.~Bielsa, S.~Li, {\it et al}, ``The BIPM Kibble balance for realizing the kilogram definition,'' \emph{Metrologia}, vol.~57, p. 045009, Jul. 2020.

\bibitem{THU2023}
S.~Li, Y.~Ma, W.~Zhao, S.~Huang,X.~Yu, ``Design of the tsinghua tabletop Kibble balance,'' \emph{IEEE Trans. Instrum. Meas.}, May. 2023.

\bibitem{NPL}
I.~A. Robinson, ``Towards the redefinition of the kilogram: a measurement of the Planck constant using the NPL Mark II watt balance,'' \emph{Metrologia}, vol.~49, no.~1, pp. 113--156, Dec. 2011.

\bibitem{NISTCS}
D.~Haddad, B.~Waltrip, and R.~Steiner, ``Low noise programmable current source for the NIST-3 and NIST-4 watt balance,'' in \emph{Proc. Conf. Precis. Electromagn. Meas. (CPEM)}, Jul. 2012, pp. 336--337.

\bibitem{haddad2016invited}
D.~Haddad, F.~Seifert, L.~Chao, {\it et al}, ``Invited article: A precise instrument to determine the Planck constant, and the future kilogram,'' \emph{Rev. Sci. Instrum.}, vol.~87, no.~6, Jun. 2016.

\bibitem{metas2022}
A.~Eichenberger, H.~Baumann, A.~Mortara, {\it et al}, ``First realisation of the kilogram with the METAS Kibble balance,'' \emph{Metrologia}, vol.~59, no.~2, p. 025008, Mar. 2022.

\bibitem{wang2014250}
N.~Wang, Z.~Zhang, B.~Han, {\it et al}, ``A 250 mA high-precision DC current source with improved stability for the joule balance at NIM,'' in \emph{Proc. Conf. Precis. Electromagn. Meas. (CPEM)}. Jul. 2014, pp. 644--645.

\bibitem{Christopher-2008}
C.~J. Erickson, M.~V. Zijll, G.~Doermann, and D.~S. Durfee, ``An ultrahigh stability, low-noise laser current driver with digital control,'' \emph{Rev. Sci. Instrum.}, vol.~79, p. 073107, Jul 2008.

\bibitem{Hall-1993}
K.~G. Libbrecht and J.~L. Hall, ``A low‐noise high‐speed diode laser current controller,'' \emph{Rev. Sci. Instrum.}, vol.~64, pp. 2133--2135, Aug 1993.

\bibitem{Daylin-2013}
D.~L. Troxel, C.~J. Erickson, and D.~S. Durfee, ``Note: Updates to an ultra-low noise laser current driver,'' \emph{Rev. Sci. Instrum.}, vol.~82, p. 096101, Sep 2013.

\bibitem{Taubman-2011}
M.~S. Taubman, ``Low-noise high-performance current controllers for quantum cascade lasers,'' \emph{Rev. Sci. Instrum.}, vol.~82, p. 064704, Jun 2011.

\bibitem{Taubman-2013}
M.~S. Taubman, ``Note: switch-mode hybrid current controllers for quantum cascade lasers,'' \emph{Rev. Sci. Instrum.}, vol.~84, p. 016103, Jun 2013.

\bibitem{Xia-2022}
D.~Yang, S.~Xia, L.~Ouyang, W.~Hou, and L.~Guo, ``An ultrahigh performance laser driver based on novel composite topology enhanced howland current source,'' \emph{Rev. Sci. Instrum.}, vol.~93, p. 123001, Dec 2022.

\bibitem{Carmine-1998}
C.~Ciof, R.~Giannetti, V.~Dattilo, and B.~Neri, ``Ultra low-noise current sources,'' \emph{IEEE Trans. Instrum. Meas.}, vol.~47, pp. 78--81, Feb 1998.

\bibitem{Scandurra-2014}
G.~Scandurra, G.~Cannatà, G.~Giusi, and C.~Ciofi, ``Programmable, very low noise current source,'' \emph{Rev. Sci. Instrum.}, vol.~47, pp. 78--81, Dec 2014.

\bibitem{Wangyanzhang-2022}
J.~Qin, Y.~Zhou, and Y.~Wang, ``A high dynamic range and ultralow-noise bipolar current source for unshielded SERF atomic magnetometers,'' \emph{IEEE Trans. Instrum. Meas.}, vol.~71, p. 2001308, Dec 2022.

\bibitem{Kyu-2005}
K.~Kyu-Tae, K.~Mun-Seog, P.~Po, Gyu, and N.~Juergen, ``Stabilization of magnet current using voltage standards,'' \emph{IEEE Trans. Magn.}, vol.~41, pp. 3760--3762, Oct 2005.

\bibitem{PTBcurrent-2019}
F.~Isaac, B.~Ralf, D.~Dietmar, {\it et al}, ``Externally referenced current source with stability down to 1 nA/A at 50 mA,'' \emph{IEEE Trans. Instrum. Meas.}, vol.~68, pp. 2129--2135, Jun 2019.

\bibitem{schlamminger2013design}
S.~Schlamminger, ``Design of the permanent-magnet system for NIST-4,'' \emph{IEEE Trans. Instrum. Meas.}, vol.~62, no.~6, pp. 1524--1530, Jun. 2013.

\bibitem{li2022irony}
S.~Li and S.~Schlamminger, ``The irony of the magnet system for Kibble balances—a review,'' \emph{Metrologia}, vol.~59, no.~2, p. 022001, Mar. 2022.

\bibitem{Stephan16}
I.~A. Robinson and S.~Schlamminger, ``The watt or \textsc{K}ibble balance: a technique for implementing the new \textsc{SI} definition of the unit of mass,'' \emph{Metrologia}, vol.~53, no.~5, pp. A46--A74, Sep. 2016.

\bibitem{li18}
S.~Li, M.~Stock, F.~Biesla, {\it et al}, ``Field analysis of a moving current-carrying coil in OMOP Kibble balances,'' in \emph{2018 International Applied Computational Electromagnetics Society Symposium (ACES)}.\hskip 1em plus 0.5em minus 0.4em\relax IEEE, 2018, pp. 1--2.

\bibitem{li2022}
S.~Li and S.~Schlamminger, ``Magnetic uncertainties for compact Kibble balances: An investigation,'' \emph{IEEE Trans. Instrum. Meas.}, vol.~71, pp.~1502409, Jul. 2022.

\bibitem{NIST2024flexures}
L.~Keck, S.~Schlamminger, R.~Theska, {\it et al}, ``Flexures for Kibble balances: Minimizing the effects of anelastic relaxation,'' \emph{Metrologia}, Jun. 2024.

\bibitem{li17}
S.~Li, F.~Bielsa, M.~Stock, {\it et al}, ``Coil-current effect in Kibble balances: analysis, measurement, and optimization,'' \emph{Metrologia}, vol.~55, no.~1, pp. 75--83, Dec. 2017.

\bibitem{nistmag}
F.~Seifert, A.~Panna, S.~Li, {\it et al}, ``Construction, measurement, shimming, and performance of the \textsc{NIST-4} magnet system,'' \emph{IEEE Trans. Instrum. Meas.}, vol.~63, no.~12, pp. 3027--3038, Jun. 2014.

\bibitem{BIPMmag2017}
S.~Li, F.~Bielsa, M.~Stock, {\it et al}, ``A permanent magnet system for Kibble balances,'' \emph{Metrologia}, vol.~54, no.~5, pp. 775--783, Sep. 2017.


\end{thebibliography}

\end{document}